%% file: 00_paper.tex
\documentclass{vgtc}                          

\graphicspath{{figures/}{pictures/}{images/}{./}} 

\usepackage{times}                     

\usepackage{tabu}                      
\usepackage{booktabs}                  
\usepackage{lipsum}                    
\usepackage{mwe}                       

\usepackage{mathptmx}                  
\usepackage{soul}
\usepackage{cancel}
\usepackage{amsmath}
\usepackage{balance}
\usepackage{amssymb}

\onlineid{0}

\vgtccategory{Research}

\vgtcinsertpkg

\title{Marks, Channels, and Dead Ends: Stop Running Graphical Perception Studies and Start Modeling Visualizations as Images}

\author{%
  \begin{tabular}[t]{c}Khairi Reda\thanks{redak@uic.edu}\\
    \parbox{1.4in}{\scriptsize\centering Electronic Visualization Lab\\University of Illinois Chicago}\end{tabular}\hspace{1em}%
  \begin{tabular}[t]{c}Shambhawi Sharma\\ 
    \parbox{1.4in}{\scriptsize\centering Electronic Visualization Lab\\University of Illinois Chicago}\end{tabular}\hspace{1em}%
  \begin{tabular}[t]{c}Luc Renambot\\
    \parbox{1.4in}{\scriptsize\centering Electronic Visualization Lab\\University of Illinois Chicago}\end{tabular}%
  \\[3em]
  \begin{tabular}[t]{c}Fabio Miranda\\
    \parbox{1.4in}{\scriptsize\centering Electronic Visualization Lab\\University of Illinois Chicago}\end{tabular}\hspace{1em}%
  \begin{tabular}[t]{c}Saeed Boorboor\\
    \parbox{1.4in}{\scriptsize\centering Electronic Visualization Lab\\University of Illinois Chicago}\end{tabular}%
}

\abstract{
    \input{sections/00_abstract}
} 


\begin{document}

\firstsection{Introduction}

\maketitle

\input{sections/01_intro}
\input{sections/02_graphical_perception}
\input{sections/03_image_models}

\input{sections/04_scatterplots}
\input{sections/05_future}
\input{sections/06_limitations}

\input{sections/07_conclusion}

\acknowledgments{This paper is based upon research supported by the US National Science Foundation under awards 2618093, 2320261, 2330565, and 2411223.}

\begingroup
\interlinepenalty=10000
\balance
\bibliographystyle{abbrv-doi}

\bibliography{0_refs}
\endgroup
\end{document}

%% file: sections/00_abstract.tex
Graphical perception studies are the visualization community's preferred tool for evaluating visualizations. By measuring how accurately people interpret various arrangements of visual marks and channels, they aim to establish best practices for encoding data visually. We argue this model of assessing visualizations is fundamentally flawed, and no amount of additional empirical studies will fix it. The problem is that visualization frameworks theorize effectiveness at the level of the encoder (i.e., which data-to-visual mappings work best for a given context). Human perception, however, consists of a fundamentally incompatible decoder that operates at a different level, namely, retinal images. This encoder-decoder asymmetry means that experimental results and guidelines are often poor predictors of actual perceptual performance. Moreover, the image that actually reaches the visual system is not determined by the encoding specification alone, but is rather emergent from interactions between encoding rules, the input data, and even micro-design parameters, all of which are invisible to encoding theory. Consequently, small changes in data distributions or seemingly minor design variations can substantially change how a visualization is perceived, even when the nominal encoding specification remains unchanged. The space of such interactions is vast and cannot be covered by running one study after another.  

We argue that visualizations should be studied as \emph{images}, and evaluated using computational models of human vision that take pixels as input. Such models capture the perceptual representations the visual system actually constructs, shifting evaluation toward modeling the decoder rather than anchoring on abstract encoding specifications. This approach is scalable, human-grounded, and sensitive to the emergent image properties that encoding theory and graphical perception studies both miss. We first describe methodological and theoretical weaknesses of the current paradigm, and propose a theory of visualization perception grounded in summary-statistical accounts of vision. We then demonstrate how image-based vision models can predict visualization discriminability in scatterplots while reproducing established results. We close by outlining a research agenda for vision-based visualization evaluation.

%% file: sections/01_intro.tex
How to design effective visualizations is arguably the central animating question of the visualization research community. Because visualizations are ultimately created for human audiences, the field has a rich tradition of evaluating them through empirical human-subject studies. In graphical perception experiments, participants are shown visualizations and asked to perform elementary perceptual tasks, such as estimating a quantity, comparing two values, and judging a trend, while the visual encoding is systematically varied. The goal is to establish which encodings let people read the data with the highest accuracy and the fewest errors. This framework began with Cleveland and McGill~\cite{cleveland1984graphical} and has continued over decades, testing an ever-increasing number of visual encodings and tasks to build what the field regards as a scientific foundation for visualization design. Systems such as Draco have also been developed to codify this empirical knowledge and operationalize it to recommend designs suited to a given context~\cite{Moritz2019-draco}.

Although important to the early development of the field, this evaluation framework is problematic on several fronts. The most obvious issue is perhaps scale, with a visualization design space that is simply enormous. The field's response has been to populate an ever-larger matrix of empirical results, documenting how each visualization technique performs under each task, one cell at a time. Given the combinatorial size of that space, it is difficult to see how the effort can be sustained. Worse, findings in one part of the matrix rarely generalize to others, as there is no principled mechanism for extrapolating from tested encodings to untested ones, so each new design combination demands new empirical studies. Moreover, this matrix is actually larger than it first appears, because effectiveness depends not only on the encoding and the task but also on the distribution of the underlying data. For example, changing how data is distributed (e.g., uniform, clustered, sparse) can drastically alter a visualization's effectiveness~\cite{xiong2022investigating}. Encodings that work well for one dataset can fail entirely for another dataset with marginally different characteristics, even when the task is identical. Accounting for this properly adds yet another dimension (i.e., Data Characteristics) to an already large empirical space.

These limitations leave perception studies with weak predictive power, and at times even contradictory accounts of how visualizations perform. Consider the case of color encodings. Based on numerous perceptual studies~\cite{kalvin2000building,rogowitz2001blair}, the field has long held that hue-based encodings (so-called `rainbow' colormaps) are a poor choice for communicating quantitative data~\cite{borland2007rainbow}, yet practitioners use them anyway~\cite{moreland2015we}. Recent studies found that rainbow encodings can be \emph{superior} to the perceptually uniform designs researchers have long recommended~\cite{reda2021color,reda2020rainbows}. Other studies, however, continue to find rainbows ineffective~\cite{liu2018somewhere,borkin2011evaluation}. So which is it? Channel rankings suggest that hue is among the least effective quantitative encodings, yet empirical evidence cited above points to conflicting results. A designer consulting the guidelines~\cite{munzner2014visualization} could thus be steered toward the best design, the worst, or possibly anywhere in between, with no way to tell short of running a new perceptual study of their own. Consequently, we are left with an encoding theory of little predictive value, and a body of empirical results that is at best hard to parse and difficult to extrapolate from.

Given the fragility of findings, it is unsurprising that many practitioners frequently set aside empirical guidance in favor of their own practice~\cite{kim2025understanding}. The exhaustive nature of perception studies is also yielding diminishing theoretical returns. It is not only impractical but arguably a poor use of the field's time to test every conceivable design, particularly when findings can be easily overturned by a subsequent study or in different tasks. To build a genuine science of visualization, we need a predictive and explanatory theory that does more than catalog performance for encoding-task pairs. 

We argue that a key problem in visualization theory is that it conflates the \emph{encoder} with the \emph{decoder}. Visualization frameworks theorize design at the level of the encoding (i.e., the formal specification that maps data to visual marks and channels), whereas the visual system operates on retinal images and constructs an entirely different perceptual representation. Because of this encoder--decoder asymmetry, properties of the encoding are a poor predictor of what the visual system will do with the image it produces. Worse, the image that the decoder receives is not specified by the encoding alone, but is rather emergent from the joint interaction of encoding rules, data distribution, and micro-design parameters, which fall beneath the radar of graphical perception and encoding theories. Two visualizations with identical encoding specifications but different data will produce qualitatively different images, and hence might be interpreted with different accuracy. Two scatterplots that differ only in whether their points are filled or outlined can differ substantially in spatial-frequency content, and thus in how the visual system represents them. By fixating on encoding choices, graphical perception studies abstract away precisely the factors that shape how a visualization is perceived. Our empirical knowledge of visualization perception is not merely incomplete but fundamentally unreliable. Fixing this requires a new theoretical framework (and accompanying methods) for analyzing visualizations.

We propose that visualizations be studied and evaluated as \emph{images}, using computational models of human vision that take pixels as input. Such models provide an account of how the decoder (i.e., the visual system) works. Modern vision models specifically seek to characterize the perceptual summary representations that the visual system constructs in response to an image~\cite{rosenholtz2012summary,freeman2011metamers,balas2009summary}. 
Evaluating visualizations at this level of analysis offers two advantages. First, by looking at images, we capture how encodings, data, and other design parameters \emph{jointly} impact the image. Second, and more fundamentally, we can ask whether the data properties a designer intends to communicate \emph{survive} the transformations imposed by the visual system, and whether they remain accessible within the perceptual system's representational space. In effect, rather than asking which mappings were used to generate a visualization, we ask what the resulting image looks like to the visual system. This approach is scalable, biologically grounded, and sensitive to exactly the emergent image properties that encoding theory and graphical perception studies both miss. As a proof of concept, we use a texture-based model of vision to show that image statistics can predict the perceptual judgments of correlation in scatterplots, reproducing established human results. We close by laying out the open research directions for this new visualization evaluation paradigm.

We should note that our critique targets graphical perception studies, where the visual encoding is the primary independent variable and accuracy or speed on perceptual tasks is the dependent measure. It is not a critique of all visualization evaluation practices, which are rich with a variety of methods~\cite{carpendale2008evaluating}. Our claim is narrower and, we think, more consequential. In that graphical perception studies are the wrong approach for determining which encodings are effective.

%% file: sections/02_graphical_perception.tex
\section{The Graphical Perception Paradigm}

The question of which visual encodings best support human perception of data has long inspired research in the visualization community. Early work in cartography and graphic semiology (e.g., by Bertin~\cite{bertin1983semiology}) grappled informally with which types of charts are most readable, but it was Cleveland and McGill's landmark study that addressed these questions using formal empirical methods~\cite{cleveland1984graphical}. Informed by method in psychophysics (in particular Stevens' power law~\cite{stevens1957psychophysical}), Cleveland and McGill decomposed visualization interpretation into a set of elementary perceptual tasks. These comprise basic perceptual judgments that a viewer must make in order to extract quantitative information from charts. Tasks tested included extracting values encoded as point positions, estimating the length of bars, or judging the relative angles between marks. By running controlled experiments in which participants made judgments across different visual encodings, they derived an empirical ranking of visual `channels' from most to least accurate. For example, they found that position along a common axis was the most accurate channel, followed by position along non-aligned axes, then length, angle, and area, with color at the bottom. This ranking became one of the fundamental empirical findings in visualization research~\cite{munzner2014visualization}, and helped establish a template for what graphical perception studies would look like for the following four decades.

Since then, the work of Cleveland and McGill has been replicated, extended, and institutionalized as the de facto framework for evaluating visual data encodings. Heer and Bostock revisited the original experiments using Mechanical Turk, demonstrating that the ranking largely held in a crowdsourced setting, making large-scale graphical perception studies feasible~\cite{heer2010crowdsourcing}. Other work has examined how visual channels perform for different data types such as ordinal versus quantitative~\cite{mackinlay1986automating}, or investigated ensemble and aggregate judgments~\cite{gleicher2013perception, mateevitsi2024science,reda2019evaluating}, the effects of visual embellishment~\cite{skau2015evaluation}, and probed a variety of conventional charts~\cite{saket2017evaluating, talbot2014four} and other lesser-known representations like horizon charts~\cite{heer2009sizing} and connected scatterplots~\cite{haroz2015connected}.

These studies have sought to systematically catalog how well people can read different charts, testing an ever-increasing number of visualization techniques in a variety of tasks (e.g., comparisons~\cite{ondov2018face}, point estimate~\cite{ware1988color}, and trend perception~\cite{robertson2008effectiveness}). The graphical perception methodology has been quite productive with hundreds of publications~\cite{quadri2021survey}. It has established important biases in visual interpretation of charts, showing, for example, that people consistently underestimate areas~\cite{flannery1971relative} or are drawn to parts of higher variation within a line chart~\cite{moritz2023average}. The methodology has also demonstrated that certain visual channels (in particular, the hue of color) are poorly suited for encoding ordered, quantitative information~\cite{kalvin2000building, rogowitz2001blair}. These findings have influenced both guidelines and have been absorbed into visualization systems, including open-source tools~\cite{harrower2003colorbrewer, wongsuphasawat2015voyager} and flagship commercial products like Tableau~\cite{mackinlay2007show}. 

\subsection{The Encoder-Decoder Asymmetry}

Despite its contribution to the field, the graphical perception paradigm rests on assumptions that are increasingly difficult to defend. The central premise is that we can predict how well people interpret visualizations by examining the encodings used to generate them. This argument makes sense under the assumption that people read visualizations by reversing the encoding by quantitatively decoding the data that was put into the image, through a process symmetric to the encoder. To read a bar chart, one remaps bar lengths back to values. To read a scatterplot, one remaps horizontal and vertical position back to x, y data dimensions. To read a color-coded scalar field, one resolves color back to the quantities it encodes. However, people rarely read charts this way. For example, seldom is anyone interested in the value of a single point in a scatterplot. What interests them instead is the \emph{emergent} pattern formed by the point cloud as a whole. The position of individual circles, while encoding the raw data, is often not of much interest. Instead, reading a scatterplot often revolves around assessing the shape and eccentricity of the point cloud~\cite{yang2018correlation}. However intuitive this task may feel, it is a fundamentally different perceptual task from assessing the position of individual marks. This means that the ranking of visual encodings is at best a poor predictor of visualization performance, because many (if not most) tasks depend not on decoding the original data, but rather on analyzing \emph{emergent features} in the resulting image.  While others have questioned the importance of channel precision as a design principle~\cite{bertini2020shouldn}, our critique cuts deeper. We argue that rankings fail to provide utility even for basic visualization uses, because the encoding specification alone does not determine the image.

\begin{figure}[t]
  \begin{center}
    \includegraphics[width=0.85\linewidth]{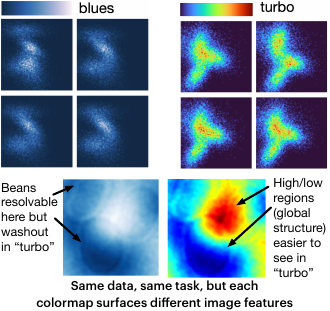}
  \end{center}
  \vspace{-5mm}
  \caption{\footnotesize \textbf{Top:} two lineups in which one of four scalar fields differs from the others. In an empirical study, a rainbow (turbo) colormap gave participants the highest chance of inferring the odd plot, whereas the perceptually uniform (blues) scale performed worst~\cite{reda2020rainbows}---a complete reversal of established guidelines. \textbf{Bottom:} On the same dataset and same odd-one-out task, the best-performing colormap itself reverses depending on what distinguishes the odd field~\cite{reda2022rainbow}. When the difference lies in a local `bean' shaped feature (most visible in the bottom-left plot), the uniform (blues) map affords the highest accuracy; when it lies in the global distribution, a rainbow colormap is the most effective choice. Essentially, the most effective colormap is dictated by emergent, data-dependent image features rather than by visual encoding guidelines or channel ranking, even when the task and data type are held the same.
  } 
 \label{fig:rainbows}
  \vspace{-4mm}
\end{figure}

The case of color-encodings makes this argument even more concrete. Years of graphical perception research have been devoted to evaluating people's ability to interpret color-coded spatial data~\cite{ware1988color}. Study after study has concluded that hue-based encodings are poorly suited for this task~\cite{moreland2015we, liu2018somewhere, rogowitz2001blair}. This is ostensibly because people cannot reliably resolve hues back to quantities, given that this channel lacks perceptual order. Consequently, so-called rainbow colormaps have long been maligned in the visualization community as the canonical example of bad visualization design~\cite{borland2007rainbow, crameri2020misuse}. Yet practitioners continue to use them, unmoved by decades of warnings from visualization researchers~\cite{ware2023rainbow}. A recent study explains this disconnect. Using an inferential task that more closely models what a scientist analyzing spatial data actually does (Figure~\ref{fig:rainbows}-Top), the study found that rainbow colormaps can in fact be superior to the perceptual scales recommended by visualization theory~\cite{reda2020rainbows}. The reason is not that hue is a good channel for recovering quantitative values (it is not~\cite{liu2018somewhere}). Rather, it is because mapping data through a hue-based colormap tends to introduce additional perceptual features into the image in the form of color categories. These manifest as bands of red, green, and blue that segment a scalar field into seemingly discrete regions. Although technically `artifacts', such features are actually quite useful: much as histogram bins help people think about univariate distributions, color bands serve as a reliable perceptual cue for analyzing spatial distributions and inferring the process that generated those distributions~\cite{reda2020rainbows}. Here, the visual encoding seems to interact with the data to imprint new features in the visualization that are entirely absent from the encoding specification. It is those features (and not the original encoding) that appear to drive the perceptual performance. 

Emergent perceptual features of this kind, like hue bands in a geographical map or a scalar field, arise from interactions between data and visual encoding rules. Yet these features are invisible to traditional channel rankings and cannot be predicted from encoding specifications alone. Whether a visualization succeeds thus depends less on which data variables are mapped to which channels, and more on the image-level structure that emerges from those interactions. The graphical perception paradigm, which is built around evaluating the encoder alone, does not offer an account of this process.

\subsection{Sensitivity to Data Distributions and Micro-Design Parameters} 

Because image-level emergent features play an outsized role, visualization perception is not just dependent on the encoding choices of the designer, but can also be highly sensitive to the distributional properties of the data being mapped~\cite{kim2018assessing}. For example, people's accuracy at estimating base rates in icon arrays depends strongly on how the icons are spatially arranged, whether randomly, diagonally, or row-wise~\cite{xiong2022investigating}. Each of these arrangements will produce systematically different estimation errors. In fact, the arrangement factor alone accounts for the majority of the variance in performance. Fortunately, spatial arrangement in the case of icon arrays is a design choice. However, for representations that rely on position channels, like scatterplots, parallel coordinates, histograms, or for spatial data where the position is given by the data itself, the designer has far fewer degrees of freedom. The effectiveness of such visualizations may thus be largely dictated by the distributional characteristics of the data being encoded, which may not be knowable at design time.

The color-encoding example above illustrates the same problem. In a second study that tested the same visual inference task (i.e., finding the plot that does not belong from a lineup of visualizations), the task favored a rainbow colormap, but only when the target visualization differed in its global distribution~\cite{reda2022rainbow}. When the difference between the plots was in local features, a uniform, single-hue color scale provided the best performance (Figure~\ref{fig:rainbows}-Bottom). In effect, subtle differences in the image dictate which visual encodings are best suited, even when the task and data type are held the same. Such data-distribution and image-level effects are rarely modeled in graphical perception studies, and more often act as confounds that quietly undermine the external validity of these studies. Accounting for these effects properly would mean adding yet another dimension to the empirical matrix (i.e., Visual Encodings $\times$ Tasks $\times$ Data Characteristics).

\begin{figure}[t]
  \begin{center}
    \includegraphics[width=\linewidth]{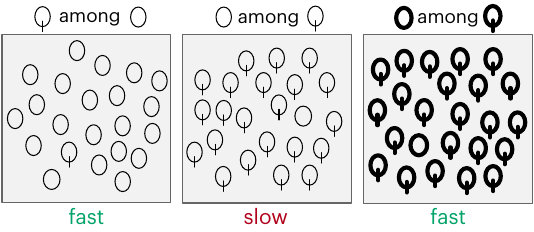}
  \end{center}
  \vspace{-5mm}
  \caption{\footnotesize Searching for a Q among Os (left) is fast, but finding an O among Qs (middle) is much more difficult~\cite{wolfe2001asymmetries}. However, if the characters are made bold (right), finding the \textbf{O} among \textbf{Qs} suddenly becomes much easier~\cite{chang2016search}. Such seemingly minor design variations can drastically affect what is easy to see in the display, and hence are likely to play an outsized (yet untheorized) role in a visualization's effectiveness.} 
 \label{fig:search_display}
  \vspace{-4mm}
\end{figure}

But it is not only data characteristics that matter. Even seemingly trivial design variations, ones that fall well beneath the threshold of what encoding theory would consider meaningful, can dramatically change what is perceptually easy or hard to see in a display. Evidence for this comes from visual search asymmetries, which are well-documented in vision science~\cite{treisman1985preattentive,wolfe2001asymmetries,thornton2007parallel}. For example, finding a \texttt{Q} in an array of \texttt{O} distractors is fast and effortless (Figure~\ref{fig:search_display}). Finding an \texttt{O} among \texttt{Qs}, on the other hand, is slow and requires serial search. However, if we were to slightly increase the stroke weight of the letters, a search for \texttt{O} among \texttt{Qs} suddenly becomes much easier again~\cite{chang2016search} (we return to this example in \S\ref{sec:PSmodel}). If a manipulation as minor as bolding can drastically alter what people can easily see in relatively simple search displays, then minor design variations in visualizations (which can involve an even wider set of micro-design choices) must be expected to produce comparably large and unpredictable effects on perception. Yet, such low-level design factors are almost never studied or controlled for. Thus, a fourth dimension should be added to the empirical matrix: Visual Encodings~$\times$ Tasks $\times$  Data Characteristics $\times$ Micro-Design Choices. Each dimension interacts with the others in ways that are difficult to anticipate. The resulting space is combinatorially large and intractable for exhaustive human-subject testing, even for a subset of visualizations. Yet, the interactions within these factors are precisely what determine whether a visualization will work in practice or not for a specific context.

\subsection{Saliency Models and CNNs in Vis Research}

One line of research has sought to model visualizations in the image space. This includes developing  \emph{visual saliency} models that predict where viewers are likely to fixate within a visualization~\cite{bylinskii2018different,janicke2010salience}. These models combine bottom-up image features (e.g., elements that pop out) with top-down priors (e.g.,  interest in text)~\cite{matzen2017data}. Yet, saliency models show weak correspondence with human fixations when viewers have been primed with a task or a message~\cite{polatsek2018exploring}. This greatly limits their usefulness for evaluating whether a visualization supports a specific goal. More recently, researchers have applied CNNs (Convolutional Neural Networks) and multimodal models to graphical perception tasks, often seeking to replicate experiments by Cleveland and McGill~\cite{haehn2018evaluating,nguyen2025evaluating}. Recent works employ computer vision and large language models to answer questions about charts~\cite{hoque2022chart} or to provide design feedback~\cite{shin2025visualizationary,kim2025automated}. While innovative, these approaches frequently diverge from human perceptual behavior and often exhibit poor generalization~\cite{cui2024generalization}. This limitation is unsurprising given that such models are optimized for predictive accuracy rather than fidelity to the visual perceptual system. Although CNNs can be a useful model of human vision, simpler architectures often exhibit better correspondence with human perceptual judgement~\cite{wallis2017parametric}, if not exact image appearance. Our proposed model also shares similarity with information-theoretic accounts, which treat the visualization pipeline as a communication channel and quantify information loss along it~\cite{chen2010information}.

\subsection {Summary}

The above analysis highlights two problems in graphical perception research: the first is methodological, and the second is theoretical. The methodological issue is one of scale, with a graphical perception enterprise built around cataloging empirical performance across a vast, combinatorial design space, which we cannot hope to cover exhaustively. The deeper problem, however, is \emph{theoretical}, in that a theory of visual encoding cannot, on its own, account for the factors that jointly influence a visualization's effectiveness. The reason is that the visual perceptual system does not operate on encoding specifications, and does not normally work to decipher channels, marks, or design abstractions. Instead, the visual system receives images of visualized data and operates entirely at the level of those images. These images are not determined by encodings alone, but rather emerge from interactions among encoding choices, data distributions, and numerous seemingly minor design parameters. Visualization encoding theory currently has no vocabulary to describe, let alone predict, such interactions. Essentially, we are left with a design theory that operates at the level of abstract encodings, and a visual system that operates at the level of images, with no bridge between the two.

This gap explains why, after forty years of graphical perception studies, we still cannot predict the perceptual effectiveness of a novel visualization design from first principles. Existing frameworks, such as the perceptual rankings of channels~\cite{munzner2014visualization}, fail to predict the effectiveness of new designs and can even yield misleading guidelines that can reverse under realistic conditions (e.g., as in the case of rainbows).

%% file: sections/03_image_models.tex
\section{Visualization Perception Belongs in Image Space} 

We argue that the image is the right level of analysis for understanding visualization effectiveness, because it is after all the actual signal that the visual system receives.  Whereas other representations of visualizations (e.g., Grammar of Graphics-style specification~\cite{wilkinson2011grammar}) reflect only an improvised proxy, the image integrates all factors that jointly determine how a visualization will be perceived, including effects of encoding choices, data distributions, and other micro-design parameters. The focus on images as a level of analysis also serves to shift visualization evaluations from chasing variations of visual encodings towards developing a model of what the decoder (i.e., the visual system) will do with the image. Thus, rather than asking whether a visualization conforms to fragile design heuristics or guidelines, \emph{we can ask a more fundamental question:} do task-relevant data features encoded in a visualization survive the transformations imposed by the human visual system, and do they remain accessible within the visual system's innate representational space? Here, a visualization succeeds not because it uses visual encodings previously deemed effective, but because the information payload carried in its image is accessible to the visual perceptual system.

To develop analytical tools for assessing visualizations at this level, we need to do two things: 1) We need to model the perceptual representations that the visual system creates in response to looking at the visualizations. 2) Once these representations are modeled, we can judge if relevant data features (patterns, summary statistics, etc) that the visualization creator intends to communicate are visible and resolvable from these representations. Modeling human vision is not a trivial challenge. Fortunately, advances in vision science provide a timely opportunity. In particular, computational models of human vision, which have been validated in natural images and texture~\cite{rosenholtz2011your,portilla2000parametric}, recently reached a level of maturity and tractability that makes them viable as predictive tools for evaluating complex stimuli like visualizations. These models offer a computational account of how visual perception arises from image statistics, which is precisely the kind of decoder-side theory that visualization research currently lacks.

\subsection{Models of Visual Perception}

The human visual system is often viewed as a massively parallel processor that transforms retinal images through cascades of filters sensitive to basic properties such as edge orientation and spatial frequency~\cite{ware2010visual}. Feature integration theory (FIT)~\cite{treisman1980feature} postulates that basic features (color, orientation, size) are processed pre-attentively and largely independently, giving rise to `pop-out' effects. This view has strongly influenced visualization encoding theory, where channels are often treated as (mostly) separably decodable information carriers~\cite{healey2011attention,munzner2014visualization}. However, FIT is now widely regarded as incomplete~\cite{humphreys2016feature,quinlan2003visual,wolfe2020forty}, as it cannot account for well-documented asymmetries in visual saliency~\cite{wolfe2001asymmetries}. Such asymmetries are especially problematic for visualization, where the spatial distribution and ratio of visual features (and hence their saliency) are largely determined by the data, over which a designer has no control.

Modern accounts of vision instead emphasize \emph{summary perceptual representations} constructed by the visual system from the joint statistics of features. There is extensive empirical evidence showing that the visual system can rapidly extract summary statistics, such as the mean size, average orientation, or overall color of an array of objects, often within 200~ms~\cite{alvarez2011representing,whitney2018ensemble,ariely2001seeing}. Rather than encoding objects individually, the visual system forms compact representations, capturing aggregate properties and central tendencies. These summary representations support rapid perception of the `gist' of scenes~\cite{oliva2006building}. This ability to very quickly extract aggregate properties is not just useful for analyzing visualized data, but may in fact be the primary way in which people `read' visualizations. As discussed, viewers rarely attempt to decode precise values for individual data points, as such data retrieval operations are better suited for tables. Instead, people often make summary judgments about the entire display, including trends, clusters, outliers, and correlations within~\cite{szafir2016four,amar2005low}.  Visualization interpretation is thus driven more by summary percepts than by pointwise decoding of channels. Such judgments are governed by aggregate statistics of features, like how tightly bunched data points are, whether their positions co-vary, or if their average is higher/lower on one side of the display.

\begin{figure}
  \begin{center}
    \includegraphics[width=1\linewidth]{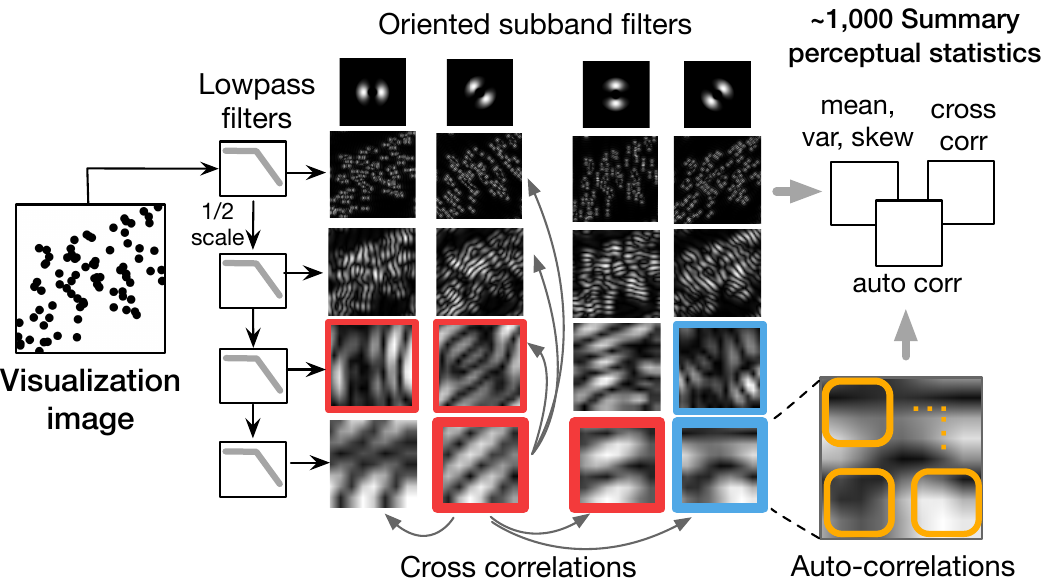}
  \end{center}
  \vspace{-5mm}
  \caption{\footnotesize Analysis of a scatterplot image through the PS model. The image is decomposed by a bank of oriented subband filters at multiple spatial scales (top: filter composition in the Fourier domain; middle: corresponding filter-response in the spatial domain). Activations are summarized (mean, skewness, autocorrelations, ...) and cross-correlated, forming a summary perceptual representation. This specific summary signature (e.g., increased/decreased mean activation in certain orientations and scales, shown here in red/blue) provides the perceptual signal for deducing a data feature of interest (x-y correlation in this example). } 
 \label{fig:pipeline}
  \vspace{-3mm}
\end{figure}

\subsection{Summary-Statistical Representations of Vision}
\label{sec:PSmodel}

One influential account of how the visual system represents summaries is based on spatial averaging of filter activations (Figure~\ref{fig:pipeline}). Specifically, responses from multi-orientation filters, which resemble complex cells in the early visual cortex, are averaged and cross-correlated within pooling regions, the size of which grows with eccentricity from the fovea. This account was formalized in the Portilla-Simoncelli (PS) model~\cite{portilla2000parametric}.
Originally developed for texture synthesis using a biologically plausible architecture~\cite{heeger1995pyramid}, the model was later shown to explain a broad range of perceptual phenomena, such as \emph{metamers} --- physically distinct images that are perceptually indistinguishable because they share the same summary statistics~\cite{freeman2011metamers,balas2006texture}. Subsequent work has shown that summary statistics also predict behavior, including the visual search asymmetries that FIT cannot explain~\cite{rosenholtz2012summary}. For example, Chang et al. explain why searching for an \texttt{O} among \texttt{Q}s (Figure~\ref{fig:search_display}) is generally difficult but becomes much easier when the characters are bolded~\cite{chang2016search}.  The reason is that thicker strokes survive the model's successive low-pass filtering; whereas the line stroke of a thin \texttt{Q} is blurred away, a bold stroke persists longer, so the cross-correlation between the circular and line features that distinguish a \texttt{Q} from an \texttt{O} is expressed across \emph{more} spatial scales. This makes the distractor \texttt{Q}s more discriminable from the target \texttt{O} than in a thin-stroke display. In addition to explaining performance in simple search, summary vision representations also predict performance in more complex visual tasks such as puzzle solving~\cite{semizer2024peripheral}, making them a potentially useful model for evaluating visualization perception.

\textbf{The PS model}, in effect, characterizes how the visual system represents images, including which features are retained (or compressed out) in the resulting perceptual summary. These summary percepts consist of about 1,000 image statistics, which are thought to correspond to correlations and information retained by the early-to-mid visual cortex, specifically in areas V1/V2~\cite{freeman2011metamers}. Figure~\ref{fig:pipeline} illustrates the architecture of this model as it reacts to a scatterplot image. The model processes the input image through a bank of wavelet filters at multiple spatial scales and orientations. Specifically, the image is first passed through a low-pass filter and convolved with a set of oriented subband filters. These Gabor-like filters act as phase-invariant edge/line detectors tuned to specific orientations, making them analogous to Complex cells in the primary cortex~\cite{martinez2003complex}. Four orientations are used in the default implementation. After filtering at a given scale, the image is downsampled by a factor of two and low-pass filtered again, with the process repeated at the coarser scale. 

The output of each filter is then summarized using a set of aggregate statistics, including mean activation, variance, skewness, kurtosis, and spatial autocorrelation. In addition to per-filter statistics, filter responses are cross-correlated with one another, both within a scale and across scales, enabling the model to capture how features at different orientations and spatial resolutions co-occur in the image. Altogether, these aggregate responses comprise on the order of a thousand statistics that jointly characterize the image. Intuitively, these statistics describe what kinds of edges are present (e.g., dominant orientations), how strongly they are expressed at different scales, where they tend to occur relative to one another, and if they repeat across spatial scales.  The model implies that if two images share the same set of summary statistics, then they are considered perceptually equivalent~\cite{wallis2017parametric}, at least to early-mid vision. 

\begin{figure}[t]
  \begin{center}
    \includegraphics[width=0.65\linewidth]{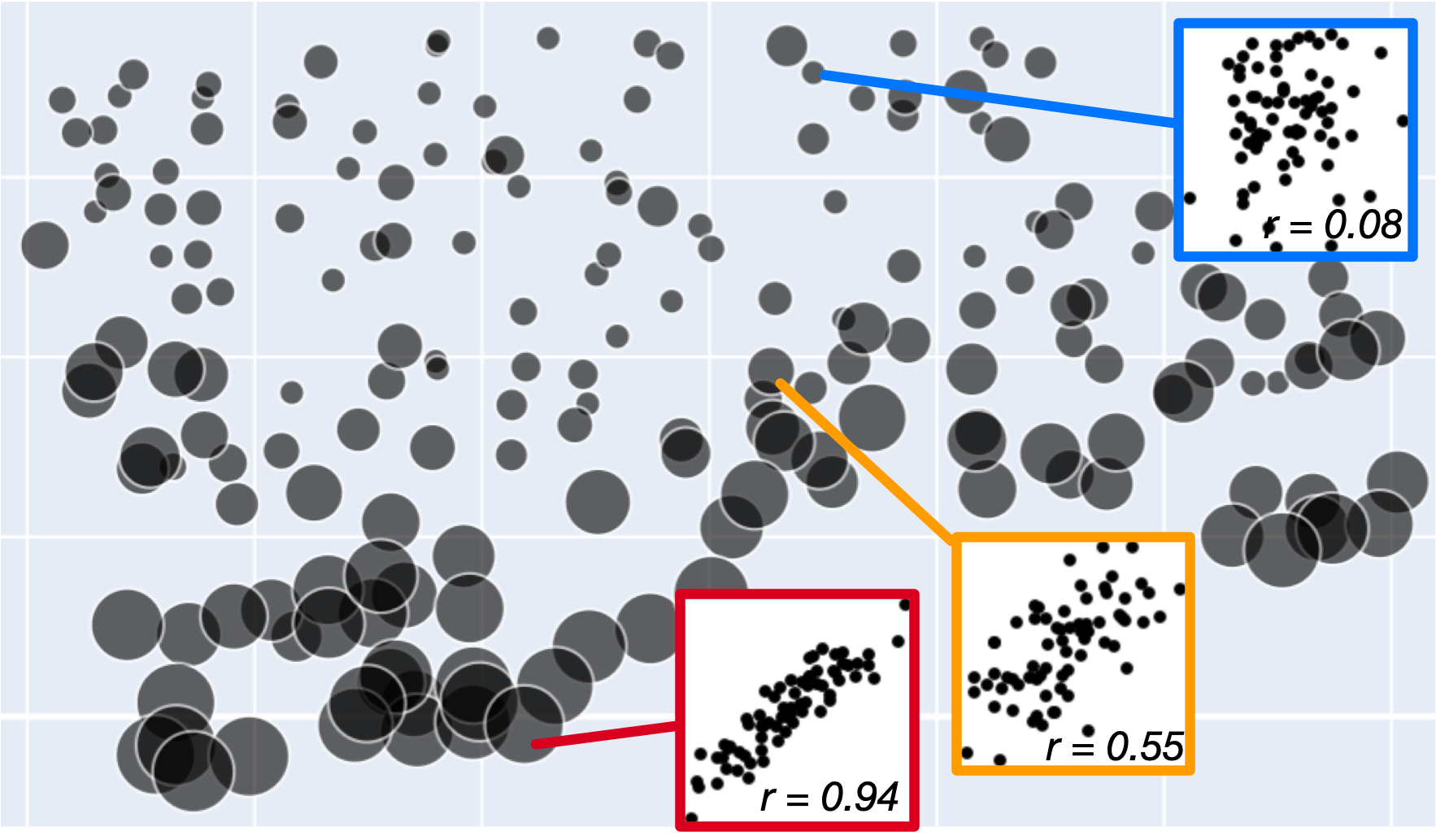}
  \end{center}
  \vspace{-5mm}
  \caption{\footnotesize Projection of 200 scatterplots based on their summary image statistics. Size indicates Pearson's correlation.} 
 \label{fig:umap}
  \vspace{-3mm}
\end{figure}

\subsection{Evaluating Visualizations through their Summary Image Statistics}
When the input image is a visualization, these summary statistics can serve as perceptual proxies for underlying data statistics. For example, co-variation in the position of marks at a specific orientation can indicate a positive correlation in a scatterplot. Likewise, whether color/luminance varies smoothly or abruptly across the image is captured by autocorrelation and cross-correlation of filter responses, which in turn can indicate the degree of spatial clustering in a choropleth map. Our \textbf{central hypothesis} is that the {extent to which summary image statistics reliably map back to the intended data features is a decisive factor for determining whether a visualization will support perception of task-relevant information}.

To illustrate, consider the scatterplot in Figure~\ref{fig:pipeline}, where the statistic of interest is Pearson's correlation. At the finest scales (rows 1 and 2), the filter responses are largely isotropic, reflecting the absence of a dominant orientation in the point cloud. As the image is progressively downsampled and low-pass filtered, individual points fuse into an oriented feature. At coarser scales, this produces increasing activation in the filter orthogonal to the direction of correlation (outlined in red). Accompanying this, we see a decrease in activation for filters aligned with the correlation diagonal (blue). This specific pattern of activations (and their cross-correlations) provides a distinctive perceptual signature that the visual system can use to infer correlation strength in scatterplots. As an example, Figure~\ref{fig:umap} shows a UMAP projection of 200 scatterplots that vary systematically in correlation ($r \in [0, 0.99]$), as characterized by their summary percepts computed through our model. As seen in this projection, image-based statistics provide an excellent basis for inferring Pearson's correlations.

%% file: sections/04_scatterplots.tex
\section{Correlation Perception Emerges from Summary Image Statistics}

To test whether summary vision models can be used to formally evaluate human performance in visualization tasks, we conducted a study on the perception of correlation in scatterplots. This is a well-studied topic in visualization research, with several works that have quantified how well people can deduce the level of correlation in scatterplots~\cite{rensink2010perception,rensink2013prospects}. We specifically use the methodology and data collected by Harrison et al.~\cite{harrison2014ranking}, where participants are shown two scatterplots side-by-side and asked to select the one that exhibits a higher correlation. We attempt to simulate this perceptual judgment using a model that takes summary image statistics as input. In effect, the model sees two scatterplot images at a time, and much like a human viewer, judges which of the two (Left vs. Right image) exhibits the higher correlation. The model only sees image statistics and does not have access to the underlying data directly. Formally, the model is described by the following classifier:

\vspace{-2mm}
\begin{equation}
\label{eq:discriminant}
f\bigl(\phi(I_1), \phi(I_2)\bigr) =
\begin{cases}
+1 & \text{if } r(D_1) > r(D_2) \\
-1 & \text{if } r(D_1) < r(D_2)
\end{cases}
\end{equation}
\vspace{-2mm}

Where $r(D_i)$ is the Pearson's correlation of a bivariate dataset $D_i$. $I=V(D)$ is the image of a dataset $D$ when seen through visualization $V$ (e.g., scatterplot). $\phi(I) \in \mathbb{R}^d$ is a \emph{d}-dimensional vector representing the $\sim$ 1,000 summary image statistics of $I$. The summary statistics are computed using the model described in \S~\ref{sec:PSmodel}. For a full description of the vision model, see~\cite{vacher2021portilla}. In effect, the classifier is designed to make an ordinal judgment based on the ground truth Pearson's correlation statistic, similar to what human subjects were asked to do. For the classifier $f$, we use a simple linear discriminant analysis (LDA) as a parsimonious model.

\subsection{Model Training}

Learning $f$ is an optimization over a loss function that counts disagreements between the classifier and the ground truth:
\begin{equation}
\mathcal{L}(f, V) =
\sum_{i < j}
\mathbb{I}\!\left[
f\bigl(\phi(V(D_i)), \phi(V(D_j))\bigr)
\;\neq\;
\operatorname{sign}\!\bigl(r(D_i) - r(D_j)\bigr)
\right]
\label{eq:loss}
\end{equation}

Where $\mathbb{I}$ is a binary indicator function. The model was trained on a series of binary trials, each comprising a pair of scatterplot images described by their summary statistics $\phi(I)$. Following the methodology of Harrison et al.~\cite{harrison2014ranking}, each trial paired a reference scatterplot generated at a fixed base correlation ($r \in \{.3, .4, .5, .6, .7, .8\}$) with a comparison scatterplot generated at varying correlation levels $[0,1]$ in steps of $.0025$, resulting in $400$ pairs. We repeat the training process five times with different randomly-seeded datasets, yielding 2000 unique stimuli per $r$ level. This ensures that the model sees examples at various levels of correlation and correlation differences.  The classifier was presented with a single difference vector $\Delta \phi = \phi(I_{Left}) - \phi(I_{Right})$, which corresponds to the offset between the image pair in the perceptual summary-statistics space. The classifier was then trained to determine, from this vector alone, which image exhibited the higher correlation. This procedure mirrored the two-alternative forced-choice paradigm of Harrison et al.

Training stimuli were generated synthetically by sampling bivariate Gaussian data at each target correlation and rendering the resulting scatterplots in the visual style of Harrison et al., essentially $100$ points drawn as filled circles of $2$-pixel radius on a $300 \times 300$ pixel canvas, without axes or labels.

\subsection{Comparing Model and Human Discriminability}

To evaluate the trained model, we simulated 100 virtual `participants'. For each virtual participant, a fresh set of test stimuli was generated using held-out random seeds not seen during training, with 400 image pairs per base $r$.
Following the methodology used for analyzing human participants, the model's binary predictions on the test pairs, along with the absolute correlation difference $\Delta r=|r_{Left}-r_{Right}|$ for each pair, were used to construct a psychometric function. Specifically, a Weibull CDF function~\cite{klein2001measuring} was fitted for each combination of virtual participant and base correlation level. From each fit, we computed just-noticeable difference (JND), which corresponds to the signal threshold (i.e., $\Delta r$ level) at which the model gives a correct response with 75\% probability (halfway between chance and perfect reliability). Note that Harrison et al. used a different convergence-based criterion to find the JNDs. To ensure model-human comparability, we refitted their binary responses to Weibull functions as well, and recomputed JNDs for human participants, which were consistent with the original results reported in the paper.

 \begin{figure}[t]
  \begin{center}
    \includegraphics[width=\linewidth]{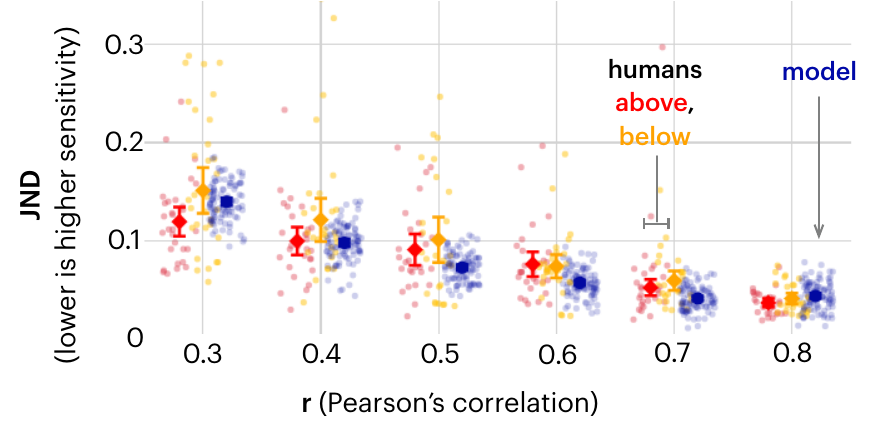}
  \end{center}
  \vspace{-5mm}
  \caption{\footnotesize JNDs from Harrison et al. (red and yellow for above and below approaches, respectively)~\cite{harrison2014ranking} versus predictions from our model (blue). A JND represents the difference in correlation between scatterplot pairs needed to make a correct judgment 75\% of the time, plotted here as a function of the correlation strength. For example, if the strength of the correlation is around $r=0.4$, the two charts need to be at about $0.1$ Pearson's units apart to be reliably discriminated.  
  Dots show JNDs for individual participants. Diamonds show means with 95\% confidence intervals (CIs). 
  Observe how the model reproduces very similar mean sensitivity results to human participants. Note also that model CIs (in blue) are narrower in part because we are able to run a larger number of `virtual' participants.  } 
 \label{fig:jnd_plot}
  \vspace{-3mm}
\end{figure}

Figure~\ref{fig:jnd_plot} shows a comparison between human and model JNDs. Lower JND implies a smaller signal is needed for discrimination, and hence better discriminability. The results of the image-based model are highly consistent with mean human thresholds. Across all six base correlations, the model produces JND estimates that fall within or immediately adjacent to the 95\% confidence intervals of the human data reported by Harrison et al.~\cite{harrison2014ranking}. A striking feature of the model results is the qualitative shape of the JND profile, which decreases monotonically as the base level of correlation increases. That is, both the model and human participants become more adept at discriminating small differences in Pearson's statistic when the correlation is higher. This effect is consistent with Weber's law, which governs a wide variety of perceptual responses. Overall, the model JNDs track closely with the human and data, reproducing the discriminability profile of a well-established result.

\subsection{Discussion}

Results show that a strictly linear discriminant operating exclusively on summary image statistics performs very closely to human-subjects' mean (if not the variance), and even exhibits Weber's law scaling. This is despite the fact that the classifier has no access to the raw data points and contains no explicit representation of chart semantics. Recall that the classifier was trained to recover the objective correlation ordering from image statistics alone, and has never observed a single human judgment. The correspondence between its JND profile and human thresholds is thus an emergent feature. That is, even though the model was optimized against ground truth (Pearson's correlation), it independently reproduced the human difficulty structure, including the sensitivity increase at higher base correlations as predicted by Weber's law.

This finding indicates that Pearson's correlation is somehow implicitly encoded in the summary image representation of a scatterplot. It also lends support to our central hypothesis: that we can tell how well people may be able to interpret a certain visualization by analyzing its image-level percepts. Notably, these percepts are the same summary image statistics that are sufficient to characterize and reproduce a wide range of textures~\cite{portilla2000parametric,balas2006texture}. Accordingly, correlation judgments (along with potentially many other perceptual inferences about a visualization) do not require channel decoding, let alone an accounting of chart specification. Instead, such judgments can be attributed solely to low-level features (and their cross-correlations) in the image space. This lends support to our thesis that visualizations should be primarily analyzed as images.

\subsection{Visualizing Same-Statistics Mongrels}

One way to get a sense of what data properties are preserved in the summary statistics is to synthesize new images that share the same statistics as the source visualization. After running the visualization through the vision model, we start from a white noise image and iteratively coerce its summary statistics so that they match those extracted from the original visualization~\cite{heeger1995pyramid,portilla2000parametric}. Because the summary percepts represent lossy compression, the resulting syntheses (referred to as ``mongrels''~\cite{balas2009summary}) will be texture-like, jumbled up images of the original visualization, potentially breaking up the semantics of the chart (e.g., by wrapping axes around or introducing illusory features). However, mongrels closely approximate the same summary statistics available to early vision, thus providing a way to intuitively probe which data statistics survive these compressive perceptual transformations. Figure~\ref{fig:mongrels} shows original scatterplots at three levels of correlation along with synthesized mongrels. The syntheses from different correlation strengths seem qualitatively distinguishable. This again suggests that correlation in scatterplots is somehow preserved in the summary statistics. Note that the mongrels are synthesized from the summary feature set $\phi(I)$ before applying the classifier $f$, so they only reflect information retained in the image statistics.

%% file: sections/05_future.tex
\section{Towards a General Visualization Decoding Model}

The above case study demonstrates how vision models, particularly those that account for the summary perceptual representation constructed by the visual system, can be used to predict visualization performance. Evaluating visualizations at the image level allows us to pursue a generalizable theory of visualization \emph{decoding}, rather than chasing the virtually endless variations of visual \emph{encodings}. This stands to shift evaluation practices away from ad-hoc, costly, and narrowly scoped user studies that can cover only a small portion of the design space, and toward principled, model-based analysis of visualizations in the image space.

To fully develop the approach, however, we must still demonstrate three things. First, the approach must extend beyond scatterplots and beyond bivariate correlation to a variety of visualization designs and tasks. Second, for a given task $T$, the model must {discriminate effective from ineffective visualizations}, by assigning better discriminability to designs that empirically support the task compared to those that do not. Third, the model must fail in human-like ways and should reproduce documented interactions, such as effects of data distributions that make it unexpectedly harder (or easier) for human observers to perform the task. 

\subsection{Extending Beyond Scatterplots}

We tested the model on scatterplots because they are well-studied and are often treated as a `fruit fly' of visualization research~\cite{rensink2013prospects}. Because the model operates on images, however, the evaluative approach should generalize to any design. Specifically, to assess whether a model supports a particular task is to essentially ask whether a task-relevant data property \emph{survives} the perceptual summary representation
imposed by the visual system. Using the loss defined in Eq.~\ref{eq:loss}, we say
that a task statistic $s(D)$ survives the perceptual bottleneck under visualization $V$ if some discriminant $f$ achieves an acceptable loss:
\vspace{-2mm}
\[
\min_{f_{T,V}}\; \mathcal{L}(f_{T,V}, V) \;\leq\; \epsilon
\]

\vspace{-3mm}
\noindent That is, if the summary statistics $\phi(V(D))$ retain enough information about $s(D)$ for a discriminant to recover the task-relevant ordering with error of at most $\epsilon$. The scatterplot case study is one instance in
which $T$ is correlation judgment, $s(D) = r(D)$ (i.e., Pearson's correlation), and $\mathcal{L}$ counts ordinal disagreements over stimulus visualization pairs.

\begin{figure}[t]
  \begin{center}
    \includegraphics[width=\linewidth]{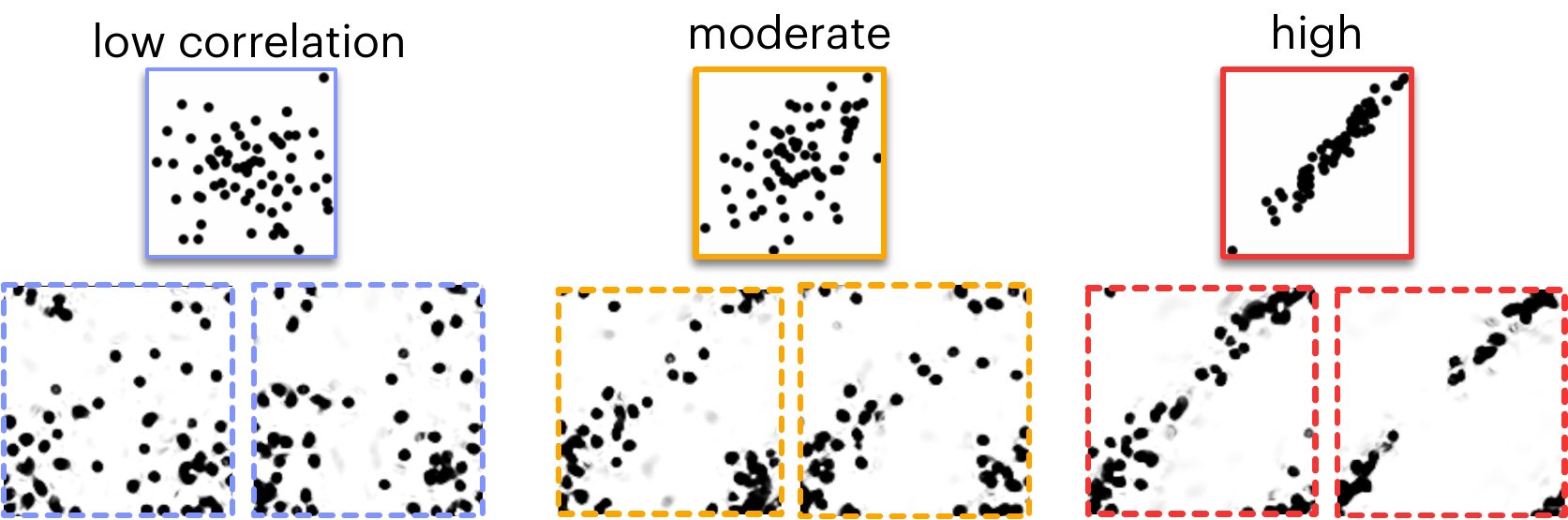}
  \end{center}
  \vspace{-5mm}
  \caption{\footnotesize Mongrels synthesized from scatterplot images of increasing correlation levels. Originals are at the top. Mongrels are generated by iteratively coercing white-noise images to match the summary statistics of the original image.  } 
 \label{fig:mongrels}
  \vspace{-4mm}
\end{figure}

Functionally, the discriminants $f_{T,V}$ are analogous to top-down visual attentional mechanisms that emphasize percepts important to the task while inhibiting others~\cite{evans2011visual,mangun1995neural,boger2021jurassic}. In effect, the discriminants will select, combine, and amplify perceptual statistics most informative for the task at hand. Therefore, $f$ must be relearned for each new combination of visualization and task statistic. This is because a new design will give rise to a different distribution of summary image statistics, and different tasks draw on various subsets of $\phi(V(D))$. For example, resolving the degree of correlation in a scatterplot relies on different image-level features than estimating the mean $y$-value of its points. However, this learning is efficient in practice and requires only 1)~identifying a task-relevant \emph{data statistic} $s(D)$, and 2)~defining a \emph{generative procedure} that produces datasets with realistic variations of that statistic. These conditions are readily met for many types of visualizations. Moreover, there exist data statistics (e.g., scagnostics~\cite{dang2014scagexplorer,wilkinson2005graph}) for describing a wide range of visual patterns and features of interest in visualizations. For cases where realistic generative models are hard to define (e.g., 3D volumes), relying on real data may be an option.

\subsection{Erring Like a Human}

A critical test for the model is whether it can discriminate effective from ineffective visualization designs. For example, scatterplots are empirically known to support accurate judgments of Pearson's correlation, whereas designs such as donut plots are not~\cite{harrison2014ranking}. The model should be sensitive to such differences. Formally, given two designs $V_{\text{good}}$ and $V_{\text{bad}}$ for a task $T$, where empirical evidence indicates that $V_{\text{good}}$ supports more accurate human judgments of statistic $s(D)$, our central hypothesis predicts:
\vspace{-2mm}
\[
\mathcal{L}\big(f_{T,V_{\text{good}}},\, V_{\text{good}}\big)
\;<\;
\mathcal{L}\big(f_{T,V_{\text{bad}}},\, V_{\text{bad}}\big)
\]

\vspace{-1mm}
Beyond an overall good vs. bad assessment of visualization, the model should predict the more unusual effects of data distribution that the literature documents but does not explain. For example, parallel coordinates are generally less effective than scatterplots, \emph{except} for weakly negative correlations, where they perform unexpectedly well~\cite{harrison2014ranking}. A well-calibrated model should reproduce this effect. This can be verified by computing model-predicted discriminability for parallel coordinates in the marginally negative correlations (e.g., $r \in [-0.25, -0.05]$). We would expect the thresholds in this $r$ range to be lower than average (higher sensitivity). Such tests can tell if the model responds to subtle, distributional effects and, critically, whether it explains them through emergent visual interactions in the image rather than through the nominal encoding.

In other words, the model and human subjects should exhibit similar modes of failure, with the model erring in the same way as humans, both across visualization designs and over variations in data distributions (e.g., random spatial arrangements in icon arrays increasing error~\cite{xiong2022investigating}). We saw some evidence of a shared error structure in the model's ability to replicate Weber's law. If this property can be demonstrated further in other designs and tasks,  it would provide even stronger evidence for a common visualization decoding theory based on summary perceptual representation. 

\subsection{Model Parameters and Degrees of Freedom}

The proposed model has several parameters and degrees of freedom that should be investigated systematically in the future. The first is the choice of the discriminant function $f$, which strongly shapes how effectively the model can exploit the available summary percepts. We adopted LDA as a parsimonious classifier with a linear decision boundary. Other alternatives include logistic regression, support vector machines (with linear and non-linear kernels), and fully connected decision layers atop the feature representation. The second is the summary feature set $\phi(I)$ itself. The current set is biologically grounded and drawn from vision research~\cite{freeman2011metamers,portilla2000parametric,rosenholtz2012summary}. However, richer representations could in principle improve correspondence with human judgments (e.g., features from deeper CNN layers),  provided they are used as a fixed perceptual front-end (e.g., borrowed from general-purpose networks like VGG~\cite{wallis2017parametric}) rather than tuned end-to-end on every task, as existing work has done~\cite{haehn2018evaluating}.

%% file: sections/06_limitations.tex
\section{Limitations}
\vspace{-1mm}
The proposed approach models only the early-to-mid stages of visual
processing, specifically the summary perceptual representation constructed in areas V1/V2. We believe that this summary image-based representation is precisely the level at which the encoder-decoder asymmetry manifests, and where graphical perception theory is currently silent. Moreover, by modeling the transformations imposed by early vision, we can capture key perceptual bottlenecks that constrain all downstream interaction with visualizations. That said, the current model has no account of working memory, attention, or higher-order cognition. Therefore, tasks that rely more heavily on these components will likely diverge from model predictions. The model thus cannot replace user studies entirely. 

There are experimental designs that could indeed probe the limits of the proposed model. Specifically, for tasks that exhibit higher divergence with human behavior, we can reduce the stimulus exposure time available to human subjects (e.g., down to 500 or even 250 ms). This serves to limit the effects of working memory and explicit reasoning, forcing viewers to rely on the percepts available to early vision. If human-model agreement improves under these shorter viewings, it would suggest that the divergence is due to post-perceptual processes that are not captured by the model. 

It is also worth noting that not all visualization studies are `graphical perception' experiments. Our critique mostly applies to the latter. More specifically, we say a study is graphical perception when the visual encoding is the primary independent variable, and accuracy (or speed) on perceptual tasks is the dependent variable. In these cases, the visual system is performing most of the inference, and performance can hence be modeled based on perceptual representations. A large fraction of the empirical visualization literature fits this description. However, there are also important visualization studies that do not. Some of those studies manipulate the \emph{viewer characteristics} rather than the encoding, for example, by measuring how performance varies with graph literacy, domain expertise, or individual traits. Other studies require conceptual comprehension of visualizations~\cite{pinker2014theory} or involve sensemaking, long-term memory, persuasion, and recall. These outcomes depend on semantics and framing, possibly more than they depend on the precise perceptual features in the visualization. For such studies, human-subject experiments remain indispensable, but they should no longer serve as the foundation for establishing encoding effectiveness. Specifically, for perceptual visualization research, where the visual encoding is the main manipulation and where tasks are largely data semantics-free, image-based vision models can be far more informative than an open-ended research program of measuring and cataloging human performance across an unbounded visualization design space.

%% file: sections/07_conclusion.tex
\vspace{-4mm}
\section{Conclusion}
\vspace{-1mm}
In this position paper, we have argued that graphical perception studies rest on a fundamental, unresolvable flaw, namely an encoder--decoder asymmetry. Whereas visualization theory analyzes designs in terms of visual encodings, the visual system operates on images that emerge from the interaction of encoding choices, data, and micro-design parameters. As a result, encodings alone are often a poor predictor of performance. We instead argue that visualizations should be studied as images, where all the factors that determine perception are integrated and the features the visual system actually uses emerge. This shifts the focus toward a first-principles model of the \emph{decoder}, i.e., what the visual system does with a visualization, rather than cataloging performance across an unbounded space of encoders. We show how a biologically inspired vision model operating on summary image statistics reproduced human judgments of scatterplots, including the sensitivity profile predicted by Weber's law.

By operating on images, a vision model of this kind is naturally applicable to many visualizations. However, validating it would require testing tasks and designs beyond scatterplots and correlation. Importantly, we would also need to demonstrate that it replicates various human failure modes. Such a model is necessarily limited to image-based percepts and cannot capture memory, attention, and higher-order cognition. Even so, we believe its value outweighs the incremental returns of graphical perception research. The field's time is better spent building general perception models than enumerating and testing endless combinations of tasks and encodings.